\RequirePackage{ifpdf}

\RequirePackage{fix-cm}

\documentclass[%
floatfix,
showkeys,
nofootinbib, %
superscriptaddress, %
]{revtex4-1}

\ifpdf
  \usepackage{cmap}
\fi
\usepackage{textcomp}

\usepackage[T1,T2A]{fontenc}

\usepackage[utf8]{inputenx}
\input{ix-utf8enc.dfu}

\usepackage[german,russian,english]{babel}

\usepackage{amsmath}
\usepackage{amssymb}

\usepackage{mathtools}
\mathtoolsset{
showonlyrefs,
mathic = true
}

\allowdisplaybreaks

\usepackage{hyperref}
\hypersetup{backref,
 colorlinks=false}
\hypersetup{pdfborder=0 0 0}

\usepackage{microtype}
\UseMicrotypeSet[protrusion]{alltext}

\usepackage{graphicx}

\usepackage[scanall]{psfrag}

\usepackage{listings}
\usepackage{listingsutf8}
\usepackage{fixltx2e}

\usepackage{nicefrac}

\makeatletter
\def\ps@pprintTitle{%
     \let\@oddhead\@empty
     \let\@evenhead\@empty
     \let\@oddfoot\@empty
     \let\@evenfoot\@oddfoot}
\makeatother

\begin{document}

\title{A practical approach to testing random number generators
in computer algebra systems}

\author{Migran N. Gevorkyan}
\email{gevorkyan-mn@rudn.ru}
\affiliation{Department of Applied Probability and Informatics,\\
  Peoples' Friendship University of Russia (RUDN University),\\
  6 Miklukho-Maklaya St, Moscow, 117198, Russian Federation}

\author{Dmitry S. Kulyabov}
\email{kulyabov-ds@rudn.ru}
\affiliation{Department of Applied Probability and Informatics,\\
  Peoples' Friendship University of Russia (RUDN University),\\
  6 Miklukho-Maklaya St, Moscow, 117198, Russian Federation}
\affiliation{Laboratory of Information Technologies\\
  Joint Institute for Nuclear Research\\
  6 Joliot-Curie, Dubna, Moscow region, 141980, Russia}

\author{Anastasia V. Demidova}
\email{demidova-av@rudn.ru}
\affiliation{Department of Applied Probability and Informatics,\\
  Peoples' Friendship University of Russia (RUDN University),\\
  6 Miklukho-Maklaya St, Moscow, 117198, Russian Federation}

\author{Anna V. Korolkova}
\email{korolkova-av@rudn.ru}
\affiliation{Department of Applied Probability and Informatics,\\
  Peoples' Friendship University of Russia (RUDN University),\\
  6 Miklukho-Maklaya St, Moscow, 117198, Russian Federation}

\begin{abstract}
This paper has a practical aim. For a long time, implementations of pseudorandom number
generators in standard libraries of programming languages had poor quality. The situation started to
improve only recently. Up to now, a large number of libraries and weakly supported mathematical
packages use outdated algorithms for random number generation. Four modern sets of statistical tests
that can be used for verifying random number generators are described. It is proposed to use command
line utilities, which makes it possible to avoid low-level programming in such languages as C or C++.
Only free open source systems are considered.
\end{abstract}


\maketitle

\section{Introduction}
\label{sec:intro}

While modeling technical systems with control it is often required to 
study characteristics of these systems. Also it is necessary to study the influence 
of system parameters on characteristics. In systems with control 
there is a parasitic phenomenon as self-oscillating mode.
We carried out 
studies to determine the region of the self-oscillations emergence. 
However, the parameters of these oscillations were not investigated. 
In this paper, we propose to use the harmonic 
linearization method for this task. This method is used in control 
theory, but this branch of mathematics rarely used in classical 
mathematical modeling. The authors offer a methodological article in order  
to introduce this method to non-specialists.

\section{Introduction}
\label{sec:intro}

Random numbers have wide applications in computer science, e.g., for
statistical testing, in cryptography, and in simulation. However, the
generation of truly random numbers is a labor-intensive task
because generators of truly random numbers are very complicated and costly. Moreover, the generation of
truly random numbers can be time consuming, and a program can be forced to wait for the next number
for a random amount of time. Most often, in computer science we deal with pseudorandom rather than
with truly random numbers. In this paper, we consider software implementations of only pseudorandom
number generators, which will be called random number generators below.

  The random number generators must satisfy the following criteria:
  \begin{itemize}
  \item to prevent cycling of the random number sequence, the period must be sufficiently long;
  \item the algorithm must be efficient in terms of time and the amount of computation resources;
  \item the algorithm must be able to reproduce the same sequence of random numbers any number of times;
  \item  the algorithm must be portable between various hardware architectures and operational environments.
  \end{itemize}

  Testing a random numbers generator is actually reduced to verifying
  that the random numbers produced by it are independent, and
  identically and uniformly distributed on the unit interval. The
  studies on this topic were performed by Soviet and Russian
  researchers.

There are various algorithm for generating random numbers (some of them are briefly described in
Section~\ref{sec:generator}), and they can have different implementations. Many theoretical studies discuss and compare
algorithms; however, we do not touch this issue and concentrate on the practical estimation of the quality
of random number sequences generated by software implementations. The estimation is performed using
software tools.

In this paper, we consider the following issues. First, open source computer algebra systems often use
random number generators that did not pass all available
tests. Second, it is preferable to use the best
(at least with respect to a certain criterion) random number generator. To resolve these issues, we describe
a structure of a test bed for investigating software implementations of random number generators. This test
bed makes it possible to estimate the current implementation of the generator used in a computer algebra
system. On addition, one can find another software implementation and build it into an open source computer algebra system.

As an illustration, we demonstrate the proposed procedure on certain computer algebra system.

\section{Generators in modern computer algebra systems}
\label{sec:generator}

The invention of the first random number generator is attributed to von Neumann in 1946. Later, in
1949, Lehmer proposed another algorithm, which was then generalized and received the name of linear
congruential generator (LCG)~\cite{knuth:aop:2::en}. This generator and its modifications became the main algorithm
implemented in the libraries of Fortran, ADA, and C.

In 1995, George Marsaglia proposed a set of statistical tests for
verifying sequences of random numbers claimed to be uniformly
distributed. The application of this test set showed that the vast
majority of random number generators produce low-quality sequences
that do not pass the majority of tests.  This set of tests became
widely known and stimulated researchers to seek better random number
generator algorithms and their implementations.

At the current time, modern versions of standard libraries in supported programming languages and
computer algebra systems, such as Maple~\cite{L_Maple}, Mathematica~\cite{L_Mathematica}, and SymPy~\cite{L_SymPy}, use implementations of
the Mersenne Twister (MT) algorithm. This algorithm received wide use as a high-quality replacement for
LCG because it is the first algorithm the implementations of which passed all tests available at that time.
This test was developed in 1997~\cite{L_Matsumoto:1998:MTE} and received its name due to the use of the Mersenne prime number
$2^{19937} - 1$. Depending on the implementation, it provides a period of up to $2^{216091} - 1$.
The main drawback of this algorithm is its awkwardness and, as a consequence, slow program code.
Note that currently much simpler and more efficient algorithms are available (see~\cite{L_xorshift:2005,L_oneill:pcg2014,L_Boldi2018}). In other
respects, this generator produces a high-quality random sequence and is applicable in the majority of
applications.

Proceed to the main goal of this paper. If a research uses the
generation of random numbers, then how can one verify the quality of a
sequence of such numbers? This can be needed if a nonstandard computer
algebra system or an outdated version is used. Even if a modern system
is used, the issue of choosing the seed value remains.

An obvious answer to this question is the use of a software package
that implements a set of statistical tests. However, all packages
known to the authors of this paper are implemented in C or C++, and
low-level programming is required to use their functions. Since
computer algebra systems use high-level problem-oriented programming
languages, the use of C and C++ functions can be difficult or even
impossible.

In our opinion, this difficulty can be overcome by using command line utilities. The use of such utilities
saves one the necessity to introduce C or C++ code in the program and replaces it by creating a script that
passes the sequence of numbers to be analyzed to the input of the testing utility.

\section{Statistical tests}

Testing random number generators is a classical problem of testing statistical hypotheses. However, in
mathematical statistics the null hypothesis should be typically refuted. By contrast, in testing random
number generators, the null hypothesis should be confirmed. In other
words, tests are designed for validating that the generated random
sequences indeed consist of independently distributed random numbers
that are not deterministically related. The generator successfully
passes the test if no statistically significant deviations from the
null hypothesis were found.

Since each random number generator is a deterministic algorithm, there always exists a statistical test
that this generator cannot pass. The generator's quality is determined by the number of tests that it can
pass. For this reason, software implementations combine tests into
test sets, which are jointly applied.

We now list some tests included in the statistical test set DieHard.
\begin{itemize}
\item The craps test. 200000 sequences of uniformly distributed random
  numbers are generated, and each of them is used to simulate the game
  of craps. Then, it is checked how well the empirical values of
  expectation and variance match their theoretical values. The
  $\chi$-squared test is used for statistical testing.
\item Birthday spacing. A random sequence on a large interval is
  generated. The spacings between the points should be asymptotically
  exponentially distributed.
\item Overlapping permutations. A large number of samples consisting
  of five consecutive random numbers are generated. The 120 possible
  orderings should occur with statistically equal probability.
\end{itemize}
DieHard includes 12 tests. Modern sets of statistical tests include
dozens of tests.

\section{Software packages of sets of statistical tests}

We have already mentioned that the first package of a set of statistical tests for random number gener-
ators was DieHard~\cite{L_DieHard} developed by G.~Marsaglia in 1995. It was distributed on a CD, and is currently
available on the Internet. Currently, DieHard is out of use, but its tests are now included in other packages.

Among the software packages containing statistical tests, we distinguish the following four.
\begin{itemize}
  
\item  TestU01~\cite{L_TestU01,L_Ecuyer:2007} developed by Pierre
  L'Ecuyer and Richard Simard. This package is written in ANSI
  C. Currently, this is the most widespread set of tests. It can test
  the generators that produce numbers in the interval $[0, 1)$. The
  latest version is 1.2.3 as of August 18, 2009.

\item PractRand~\cite{L_PractRand} developed by Chris
  Doty-Humphrey. This package is written in С++11 with elements of
  C99. It takes a stream of bytes at its input and is able to test 32-
  and 64-bit generators, and it can cope with large amounts of
  data. The latest version is 0.94 as of August 4, 2018.

\item gjrand~\cite{L_gjrand}. There is no author’s name on the
  official site of this package. It is written in С99. Takes a stream
  of bytes at its input and is distributed with a set of various
  generators that are able to produce not only uniformly distributed
  sequences but also normal, Poisson’s, and some other
  distributions. The latest version is 4.2.1 as of November 28, 2014.

\item DieHarder~\cite{L_DieHarder} developed by R. Brown. It is
  declared to be as a successor of DieHard. It is written in C and
  requires the library GSL~\cite{L_GSL} for its operation; it is able
  to test any generator that has the interface like the interfaces of
  the generators included in GSL. The latest version is 3.31.1 as of
  June 19, 2017.
\end{itemize}

Table~\ref{table:tests} lists the basic characteristics of the packages
discussed in this paper. The column \emph{Unix} indicates if the
package can be installed under *nix systems. The column \emph{Windows}
has the sign plus if the program can be built without installing
CygWin or MinGW. By expending sufficient effort, each of these
libraries can be compiled under Windows as well.

\begin{table}
  \caption{Summary characteristics of testing packages}
  \label{table:tests}
  \centering
  \begin{tabular}{| c || c | c | c | c | c | c | c |}
    \hline
    \textbf{Пакет} & \textbf{Язык} & \textbf{CMD} & \textbf{Unix} & \textbf{Windows} & \textbf{Версия} & \textbf{Год} & \textbf{Сайт} \\ \hline
    \texttt{TestU01} & ANSI C & - & ++ & $\pm$ & 1.2.3 & 18.08.2009 & \cite{L_TestU01}\\ 
    \texttt{PractRand} & C99, C++11 & + & + & + & 0.94 & 04.08.2018 & \cite{L_PractRand}\\
    \texttt{gjrand} & C99 & + & + & $\pm$ & 4.2.1 & 28.11.2014 & \cite{L_gjrand}\\
    \texttt{DieHarder} & C99 & + & ++ & $\pm$ & 3.31.1 & 19.06.2017 & \cite{L_DieHarder}\\\hline
  \end{tabular}
\end{table}

All the test packages listed in Table~~\ref{table:tests} have open source code. TestU01 and DieHarder are available for
installation in the official repositories of distribution kits, in particular Ubuntu 18.10. The column CMD
indicates the presence or absence of the command line utility. The double plus in the column Unix marks
the program packages included in the repositories. The two other packages are installed by compiling the
source codes. The symbol $\pm$ in the column Windows marks the packages the compilation of which
requires the emulators CygWin or MinGW. Each package can be used by linking the library to a С or С++
program and through the command line utility (except for TestU01). DieHarder has the richest command
line utility.

\subsection{Installation of packages under Unix-type OS}

Here we describe the installation of the packages into the user's home directory without administrator
rights. The installation was performed under GNU/Linux Ubuntu 18.10. The set of compilers gcc
version 8 was used. It is seen from the documentation to the packages that any compiler supporting the C
standards up to C99, inclusive, and C++ up to С++11, inclusive, will do.

In the user’s home directory, create the following hierarchy of directories:
\begin{lstlisting}
  mkdir -p ~/usr/bin ~/usr/lib ~/usr/share ~/usr/include
\end{lstlisting}

\begin{itemize}
\item  The directory \path{~/usr/bin} for executable files.
\item  The directory \path{~/usr/lib} for shared and static libraries (.so and .a).
\item  The directory \path{~/usr/include} for header files.
\item  The directory \path{~/usr/share} for examples and documentation.
\end{itemize}
To the file \path{~/.bashrc}, we add the following environment variables:
\begin{lstlisting}
  export PATH="$HOME/usr/bin/:$PATH"
  export LD_LIBRARY_PATH="$HOME/usr/lib:$LD_LIBRARY_PATH"
  export LIBRARY_PATH="$HOME/usr/lib:$LIBRARY_PATH"
  export C_INCLUDE_PATH="$HOME/usr/include:$C_INCLUDE_PATH"
\end{lstlisting}
This makes it possible for the command interpreter to seek the executable files also in the directory
\path{~/usr/bin} and allows the compiler to automatically include libraries and header files.

\subsubsection{Installation of TestU01}

To install TestU01, download the zip archive from the official site~\cite{L_TestU01},
unzip it, and go to the root directory:
\begin{lstlisting}
  wget http://simul.iro.umontreal.ca/testu01/TestU01.zip
  uz TestU01.zip
  cd TestU01-1.2.3/
\end{lstlisting}
TestU01 is distributed with the set of scripts Autoconf and Automake; therefore, the compilation
and installation processes are reduced to the execution of the three following commands:
\begin{lstlisting}
  ./configure --prefix=$HOME/usr
  make
  make install
\end{lstlisting}
The option \verb|--prefix=$HOME/usr| makes it possible to install the
program locally into \path{~/usr}.

After the installation, the directory \path{~/usr/include} will
contain a lot of header files. To better organize them, we create the
directory \path{~/usr/include/testu01} and move into it all \verb|.h|
files of TestU01.  The same can be done in the directory
\path{~/usr/lib} containing library files.

\subsubsection{Installation of gjrand}

Download the source codes archive from the official site~\cite{L_gjrand}:
\begin{lstlisting}
  wget https://datapacket.dl.sourceforge.net/project/gjrand/gjrand/ gjrand.4.2.1/gjrand.4.2.1.tar.bz2
  tar -xvjf gjrand.4.2.1.tar.bz2
  cd gjrand.4.2.1
\end{lstlisting}
\verb|gjrand| is compiled using the bash-scripts \verb|compile|. The
library source files are in the directory \path{src}:
\begin{lstlisting}
  cd src && ./compile
\end{lstlisting}
At the output, we obtain the compiled files of the dynamic
\verb|gjrand.so| and static \verb|gjrand.a| libraries, which we
manually move to the directory \path{~/usr/lib}:
\begin{lstlisting}
  cp -t ~/usr/lib/gjrand gjrand.a gjrand.so
  cp -t ~/usr/include/gjrand gjrand.h
\end{lstlisting}
Return to the root directory using the command \verb|cd ..| and go through all other subdirectories in the
same way. Each of them contains the script \verb|compile|, which must be executed:
\begin{lstlisting}
  cd testother/src && ./compile
\end{lstlisting}
At the output, we obtain the executable files in \path{testother/bin}. Again, return one level up \verb|cd ..|,
go to the following directory, and execute
\begin{lstlisting}
  cd testmisc/ && ./compile
\end{lstlisting}
to obtain the executable file \verb|kat|, which can be used to check the validity of the library operation. This
utility performs a number of tests to check if the program works as designed.

Go to the following directory and execute
\begin{lstlisting}
  cd testunif/src && ./compile
\end{lstlisting}
to obtain at the output the executable files for each static test. They are placed into the directory
\path{testunif/bin}. Two files \verb|mcp| and \verb|pmcp| will appear in \path{testunif}; they are the command line utilities
for testing the generated bit sequences. These utilities use executable files residing in the directory
\path{testunif/bin}; therefore, they cannot be placed into another directory.

The directory
\begin{lstlisting}
  cd testfunif/src && ./compile
\end{lstlisting}
is completely similar to the preceding directory, but the tests are intended for \verb|double numbers| in the
interval $[0, 1)$.

The directory \path{testother}
\begin{lstlisting}
  cd testother/src && ./compile
\end{lstlisting}
contains tests for nonuniform distributions. The executable files will also be placed into \path{testother/bin}.

As a result of installing \verb|gjrand|, we obtained two library files in the directory \path{~/usr/lib/gjrand}
and a header file in the directory \path{~/usr/include/gjrand}. The other utilities remain in the corresponding directories.

Note that the installation completed without errors. The package author used the compiler option \verb|-Wall|,
and only insignificant warnings about the use of the \verb|if| without parentheses were obtained.

\subsubsection{Installation of DieHarder}

To build the program, the library GSL (GNU Scientific Library)~\cite{L_GSL}
is needed. In Ubuntu, it can be installed by executing the command
\begin{lstlisting}
  apt install libgsl-dev
\end{lstlisting}

Download the source code archive from the official site~\cite{L_DieHarder} and unzip it:
\begin{lstlisting}
  wget https://webhome.phy.duke.edu/~rgb/General/ dieharder/dieharder-3.31.1.tgz
  uz dieharder-3.31.1.tgz
  cd dieharder-3.31.1
\end{lstlisting}

To install DieHarder, execute as in the other cases
\begin{lstlisting}
  ./configure --prefix=$HOME/usr
  make
  make install
\end{lstlisting}
Additionally, when \verb|./configure| is executed, the option \verb|--disable-shared| may be used,
which results in the static compilation of the command line utility, and no dynamic library will be created.
This is convenient if only the command line utility is needed.

In the process of compiling, there was an error---the compiler could not find the declaration of the type
\verb|intptr_t|. This can be repaired by including the header file \verb|stdint.h| into the file
\path{./include/dieharder/libdieharde.h}. The further installation completed without errors.
While \verb|make install| was executed, the library files were moved into the directory \path{~/usr/lib}, the
subdirectory \path{dieharder} was automatically created in \path{~/usr/include}, and all header files was
placed into it. The command line utility \verb|dieharder| was also placed into \path{~/usr/bin}.

\subsubsection{Installation of PractRand}

Download the source code archive from the official site~\cite{L_PractRand} and unzip it:
\begin{lstlisting}
  wget https://netcologne.dl.sourceforge.net/project/ pracrand/PractRand_0.94.zip
  uz PractRand_0.94.zip
\end{lstlisting}

The archive contains the ready to use dynamic and static libraries for
Windows. There is also the project file for Visual Studio. For the
installation under Unix, go to the directory \path{unix} and start the
building process using the command \verb|make|:
\begin{lstlisting}
  cd PractRand_0.94/unix
  make
\end{lstlisting}
This requires a compiler for C++ that supports the C++11 standard.

During the building process, there were errors related to letter cases in header file names. For example,
the file \verb|Coup16.h| was mentioned in \verb|test.cpp| in lowercase letters even thought the file name begins
with an uppercase letter. The same error was found in the file names \verb|NearSeq.h|, \verb|birthday.h|, and
the directory \path{Tests}. Probably, this is because the author developed the program under Windows, in
which file names are case insensitive.

After eliminating the errors, we obtain four executable files and the
static library \path{libpracrand.a}.  We have described the installation of
four program packages, and now we describe the utilities for testing
sequences of random numbers provided by these packages.

\subsection{Command line utilities}

The packages \verb|PractRand|, \verb|DieHarder|, and \verb|gjrand| provide command line utilities. These utilities
make it possible to execute statistical tests on a sequence of random numbers read from the standard input
stream or from a file. The most functionally rich utility is included in \verb|DieHarder|.

\subsubsection{PractRand utilities}

After compiling and building PractRand, we obtain four executable files:
\begin{itemize}
\item \verb|RNG_output| runs one of the generators included in the package.
\item \verb|RNG_test| is the utility designed for testing the generators distributed with the package or the data
stream obtained from the standard input.
\item \verb|RNG_benchmark| measures the performance of embedded generators and the generators added by
the user.
\item \verb|Test_calibration| is used for testing and configuring test sets.
\end{itemize}

Testing is performed using \verb|RNG_test|. For example, to run the test for the embedded generator
\verb|jsf32|, it is sufficient to execute the command
\begin{lstlisting}
  RNG_test jsf32
\end{lstlisting}
For testing an external program, one should feed the generated binary stream of unsigned integers to
the standard input of \verb|RNG_test|, e.g.,
\begin{lstlisting}
  ./random -G 4 -N 1000000000000000 | RNG_test stdin32
\end{lstlisting}
The option \verb|stdin32| makes \verb|RNG_test| to interpret the
stream of binary data as a set of 32-bit numbers. If the option
\verb|stdin64| is indicated, then the numbers are interpreted as 64-bit ones;
with the option
\verb|stdin|, the program decides for itself how to interpret the input
data. For testing data from a file, the following command can be used:
\begin{lstlisting}
  cat file.data | RNG_test stdin64
\end{lstlisting}
The data in the file must be binary; text files are not supported.

The tests are executed fairly quickly, and the user gets a report at the output in which the failed and
suspicious tests are listed. With the option \verb|-p1|, the program prints all results (including the passed tests).

\subsubsection{gjrand utilities}

After building, a lot of executable file are obtained. Two of them are designed for
testing external generators. One of them resides in the directory \path{testfunif} and is called \verb|mcp| (master
computer program); the second one is in the directory \path{testother} and is called \verb|fmcp|. The utility \verb|mcp|
is designed for testing sequences of unsigned random numbers, and \verb|fmcp| tests numbers in the interval $[0, 1)$.
In other respects, there are no differences between them.

\begin{table}
  \caption{Options mcp of the package gjrand}
  \label{table:gjrand}
  \centering
  \begin{tabular}{| l | l |}
    \hline
    \textbf{Option} & \textbf{Description} \\ \hline
    \texttt{--tiny} & (10 MB)  \\
    \texttt{--small} & (100 MB) \\ 
    \texttt{--standard} & (1 GB) (default) \\
    \texttt{--big} & (10 GB) \\ 
    \texttt{--huge} & (100 GB)\\
    \texttt{--tera} & (1 TB) \\
    \texttt{--ten-tera} & (10 TB) \\
    \texttt{number} & Number of bytes \\
    \texttt{--no-rewind} & Do not rewind \\
    \hline
  \end{tabular}
\end{table}

The program \verb|mcp| takes only the binary stream at the standard input. This is done using the pipe
\begin{lstlisting}
  ./random -G 4 -N 1000000000000000 | mcp --big
\end{lstlisting}
The input stream is subjected to all set of tests. The results are printed as they are produced. The pro-
gram has no other options, except for the amount of data to be tested (see Table~\ref{table:gjrand}).

\subsubsection{dieharder utility}
\label{dieharder}

After the package \verb|DieHarder| has been installed, the command line utility
\verb|dieharder| becomes available. This utility can run and test the
library \verb|GSL| and the random number generators included in \verb|DieHarder|. In
addition, it can test streams of
random numbers loaded from files or
from the standard input stream. A detailed help on all available options can be obtained by executing the
utility with the option \verb|-h|. Here we list the most important options:
\begin{itemize}
\item  \verb|-l| show the list of available tests,
\item  \verb|-dn| select the test n for application,
\item  \verb|-a| apply all tests,
\item  \verb|-g -1| show the list of available random number generators. The selected generator can be tested by
specifying the option \verb|-g n|, where \verb|n| is the index of the
generator in the list.
\end{itemize}

In particular, the generators have the following options:
\begin{itemize}
\item  The option \verb|200| instructs the utility to take the stream of binary data fed to the standard input stdin.
\item  The options \verb|201| and \verb|202| switches on reading data from a binary and text file, respectively.
\item  The options \verb|500| and \verb|501| make it possible to read the stream of bytes from the pseudodevices
\verb|/dev/random| and \verb|/dev/urandom|.
\end{itemize}

For testing external generators, the preferable method is to pass a continuous stream of binary data
through the pipe (\verb+|+). For example, this can be done as follows:
\begin{lstlisting}
  ./random -G 4 -N 1000000000000000 | dieharder -g 200 -a
\end{lstlisting}
The program \verb|random| uses the generator number 4 for generating $10^{15}$ numbers. Note that this does
not cause the process hangup because \verb|dieharder| will close the channel and terminate the process after
completing all tests.

In the case of reading random numbers from a text file, this file must have a specific structure. Each
random number must be on a separate line, and the first lines of the file must contain the following data:
type of data (\verb|d| indicates that the numbers are long integers), the number of numbers in the file, and the
number of bits in them (32 or 64 bits). Here is an example of such a file:
\begin{lstlisting}
    type: d
    count: 5
    numbit: 64
    1343742658553450546
    16329942027498366702
    3111285719358198731
    2966160837142136004
    17179712607770735227
\end{lstlisting}
As soon as such a file is created, it can be passed to \verb|dieharder| for testing:
\begin{lstlisting}
dieharder -a -g 202 -f file.in > file.out
\end{lstlisting}
The flag \verb|-f| specifies the input file containing the numbers for the analysis. The testing results will be
saved to \verb|file.out|. For a full-fledged test, more than $10^9$ numbers are required; therefore, the file size
can be as large as tens of gigabytes or more. If the number of numbers is less, then \verb|dieharder| can begin
to read the file from the beginning, which can deteriorate the testing results.

\section{Example of testing the generators SymPy and Maxima}

Let us discuss how the testing of an arbitrary random number generator can be organized using the
computer algebra system Maxima and the library SymPy for Python. We will use CAS Maxima version 5.42.1
and Python 3.6.8, the installer Miniconda. We do not want to compare various implementations of gen-
erations; rather, we provide an example that can be used for testing generators in other libraries, packages,
and computer algebra systems.

Since SymPy is a python module, we use the standard module \verb|random| for
generating random numbers. The following code fragment outputs
unsigned 64-bit random integers into the standard output
stream.

\begin{lstlisting}
import random
import sys
while True:
    r = random.randint(0,2**64-1)
    r = r.to_bytes(8, byteorder="little", signed=False)
    sys.stdout.buffer.write(r)
\end{lstlisting}

This fragment generates an integer in the interval $[0, 2^{64})$, which is then transformed to the binary form
and placed into the standard output stream. By placing this script into the file \verb|rand_test.py|, we can
organize testing by executing the command line instruction
\begin{lstlisting}
  python rand_test.py | dieharder -g 200 -a
\end{lstlisting}
or
\begin{lstlisting}
  python rand_test.py | RNG_test stdin64
\end{lstlisting}

In the case of Maxima, a sequence of random integers can be generated using the code fragment
\begin{lstlisting}
  for i: 1 thru 10 step 1 do print(random(2^64-1))$
\end{lstlisting}
However, at the output we obtain a column of integers in text format; for this reason, we use the ability
of \verb|dieharder| to read text data from files (see Subsection~\ref{dieharder}).

Since both computer algebra systems use the Mersenne twister
algorithm, the testing results are almost identical. The utility
\verb|RNG_test| for SymPy produces the result
\verb|no anomalies in 133 test result(s)|, while DieHarder shows in
the report five weakly passed tests out of more than thirty. The
results produced by DieHarder for Maxima are almost the same.

\section{Conclusion}
\label{sec:conclusion}

A detailed description of the experience in building a test bed for
investigating software implementations of pseudorandom number
generators is given. The software described in this paper provides a
researcher with a selection of three command line utilities. These utilities eliminate the need for low-level
programming. They can be conveniently used with an intermediate connecting programming language.

To use these utilities, one should place the generated random numbers
into the standard output stream in binary form. Since the majority of
computer algebra systems are designed for high-level programming, it
can be difficult to provide data in binary form. In this case, only
the utility \verb|dieharder| can be used because it can accept data in text
format.

At the end of the paper, methods of investigating software
implementations of random number generators are demonstrated using two
open source computer algebra systems as examples.

\begin{acknowledgments}

The publication has been prepared with the support of the ``RUDN University Program 5-100''.

\end{acknowledgments}

  \bibliography{rnd-cas}

\end{document}


\title{Практический подход к тестированию генераторов случайных
    чисел систем компьютерной алгебры}

\author{М. Н. Геворкян}
\email{gevorkyan-mn@rudn.ru}
\affiliation{Кафедра прикладной информатики и теории вероятностей,\\
  Российский университет дружбы народов,\\
  117198, Москва, ул. Миклухо-Маклая, д. 6}

\author{Д. С. Кулябов}
\email{kulyabov-ds@rudn.ru}
\affiliation{Кафедра прикладной информатики и теории вероятностей,\\
  Российский университет дружбы народов,\\
  117198, Москва, ул. Миклухо-Маклая, д. 6}
\affiliation{Лаборатория информационных технологий,\\
  Объединённый институт ядерных исследований,\\
  ул. Жолио-Кюри 6, Дубна, Московская область, Россия, 141980}

\author{А. В. Демидова}
\email{demidova-av@rudn.ru}
\affiliation{Кафедра прикладной информатики и теории вероятностей,\\
  Российский университет дружбы народов,\\
  117198, Москва, ул. Миклухо-Маклая, д. 6}

\author{А. В. Королькова}
\email{korolkova-av@rudn.ru}
\affiliation{Кафедра прикладной информатики и теории вероятностей,\\
  Российский университет дружбы народов,\\
  117198, Москва, ул. Миклухо-Маклая, д. 6}

\begin{abstract}

  Данная работа носит практических характер. Долгое время реализации
  генераторов последовательностей псевдослучайных чисел в стандартных
  библиотеках языков программирования и математических пакетов были
  плохо проработаны. Ситуация начала улучшатся сравнительно
  недавно. До сих пор большое количество библиотек и слабо
  поддерживаемых математических пакетов используют в своем составе
  старые алгоритмы генерации псевдослучайных чисел. Мы описываем
  четыре актуальных набора статистических тестов, которые можно
  применить для проверки генератора, который используется в той или
  иной программной системе.  В работе предлагается использовать для
  исследования утилиты командной строки, что позволяет избежать
  низкоуровневого программирования на языках типа С или С++. Кроме
  того, рассматриваются только свободные системы с открытым
  программным кодом.
\end{abstract}

  \keywords{генерация псевдослучайных чисел, TestU01, PractRand,
    DieHarder, gjrand}

\maketitle

\section{Введение}
\label{sec:intro}

При моделировании технических систем с управлением возникает
необходимость исследования их характеристик.
Также необходимо исследование влияния параметров систем на эти
характеристики.
В системах с управлением возникает такое паразитное явление, как
автоколебательный режим.
Нами проводились исследования по определению области
возникновения автоколебаний. Однако параметры этих автоколебаний нами
не исследовались. В данной статье мы предлагаем использовать метод
гармонической линеаризации для данной задачи. Этот метод применяется в
теории управления, однако данный раздел математики достаточно редко
используется в классическом математическом моделировании. Авторы
предлагаю методическую статью, призванную познакомить неспециалистов с
применением этого метода.

\section{Введение}
\label{sec:intro}

В информатике и вычислительной технике случайные числа находят широкое
применение: для проведения статистических испытаний, в криптографии, в
имитационном моделировании. Однако получение истинно случайных чисел
является крайне трудоёмким процессом. Эта вызвано в первую очередь
сложностью и дороговизной генераторов истинно случайных чисел. Также
следует учесть, что генерация истинно случайных чисел может занимать
много времени, и вполне вероятен вариант, когда при исчерпании
источника истинно случайных чисел программа переходит в режим ожидания
(время которого тоже случайно).  Таким образом чаще речь идёт не об
истинно случайных числах, а о псевдослучайных числах. В данной работе
мы будем рассматривать программные реализации только генераторов
псевдослучайных чисел.

Генераторы псевдослучайных чисел должны удовлетворять ряду критериев~\cite{drozdova:2017:rn-generator}:
\begin{itemize}
\item для предотвращения зацикливания
  последовательности псевдослучайных чисел необходимо иметь достаточно
  длинный период;
\item алгоритм должен быть эффективным по скорости работы алгоритма и
  затрате вычислительных ресурсов;
\item алгоритм должен удовлетворять критерию воспроизводимости, то
  есть должна быть возможность воспроизвести ранее сгенерированную
  последовательность псевдослучайных чисел любое количество раз;
\item алгоритм должен быть переносим по отношению к архитектурам
  оборудования и операционным окружениям;
\item алгоритм должен быть быстрым и ресурсосберегающим.
\end{itemize}

Тестирование генераторов псевдослучайных чисел фактически заключается
в проверке последовательности псевдослучайных величин на независимость,
одинаковую распределённость, равномерность на единичном интервале.
Работы по данной тематике активно велись и отечественными
учёными~\cite{kolchin:book:random-variations,tikhomirova:2006:statistical-criteria,tikhomirova:2010:empty-box,kirichenko:2009:rng-compare}.
Также доказательство статистической выявляемости псевдослучайных чисел
можно найти, например, в работах
Ю.~Н.~Тюрина~\cite{tyurin:1998:stat-analysis}.

В основе каждой программной реализации генератора псевдослучайных чисел
лежит свой алгоритм. Во многих теоретических исследованиях обсуждаются
и сравниваются сами алгоритмы, но мы оставим данный аспект без
обсуждения. В работе мы сосредоточимся на вопросах практической
проверки качества генерируемых программным кодом последовательностей
псевдослучайных чисел. Исследование проводится с помощью программных
инструментов.

В рамках работы мы затрагиваем следующие проблемы. Во-первых, в
свободных системах компьютерной алгебры зачастую используют генераторы
псевдослучайных чисел, не прошедшие всех возможных тестов. Во-вторых,
желательно использовать наилучший (хотя бы по некоторым критериям)
генератор псевдослучайных чисел.  Для решения данных проблем нами
описывается практическая структура стенда для исследования программных
реализаций генераторов псевдослучайных чисел. Это позволяет оценить
текущую реализацию генератора, встроенную в ту или иную систему
компьютерной алгебры. Кроме того, можно подобрать другую программную
реализацию для встраивания её в свободную систему компьютерной
алгебры.

В качестве иллюстрации методика исследования продемонстрирована на
некоторых системах компьютерной алгебры.

\section{Генераторы в современных системах компьютерной алгебры}
\label{sec:generator}

Изобретение первого алгоритма генерации последовательности
псевдослучайных чисел приписывают Дж. фон Нейману в 1946 году. После
этого в 1949 году Д. Г. Лехмер (D. H. Lehmer) предложил свой алгоритм,
который впоследствии был обобщен и стал известен как линейный
конгруэнтный генератор (linear congruential generator~---
LCG)~\cite{knuth:aop:2::ru}. Именно этот генератор в различных его
модификациях стал основным алгоритмом, реализованным в библиотеках на
языках Fortran, ADA, C.

В 1995 году Джордж Марсальея (George Marsaglia) выпустил набор
статистических тестов, который позволял проверить насколько случайную
и равномерно-распределенную последовательность чисел дает тот или иной
алгоритм. Применение данного набора тестов показало, что подавляющее
большинство генераторов случайных чисел на практике дают
некачественную последовательность и проваливают большинство тестов.

Данный набор тестов пробрел широкую известность и побудил специалистов
начать поиски более качественных алгоритмов генерации случайных чисел
и их программных реализаций.

К настоящему моменту в современных версиях стандартных библиотек
активно поддерживаемых языков программирования и систем компьютерной
алгебры, таких как Maple~\cite{L_Maple},
Mathematica~\cite{L_Mathematica}, SymPy~\cite{L_SymPy} применяется
реализация алгоритма вихрь Мерсенна (MT~--- \textbf{M}ersenne
\textbf{T}wister). Именно этот алгоритм стали повсеместно внедрять как
качественную замену LCG, так как он был одним из первых алгоритмов,
программная реализация который проходил все имеющиеся на тот момент
тесты. Он был разработан в 1997 году~\cite{L_Matsumoto:1998:MTE} и
получил свое название из-за использования простого числа Мерсенна
$2^{19937} - 1$. В зависимости от реализации он обеспечивает период
вплоть до $2^{216091} - 1$.

Основным недостатком алгоритма является относительная громоздкость и,
как следствие, сравнительно медленная работа программного
кода. Заметим также, что в настоящее время разработаны намного более
эффективные и простые
алгоритмы~\cite{L_xorshift:2005,L_oneill:pcg2014,L_Boldi2018}. В
остальном же данный генератор обеспечивает псевдослучайную
последовательность хорошего качества и вполне применим для большинства
задач.

Перейдём, однако, к основной цели данной работы. Если в
исследовательской работе используется генерация псевдослучайных чисел,
то какими средствами можно проверить качество последовательностей этих
чисел? Это может быть актуально при использование нестандартной
системы компьютерной алгебры или системы старых версий. Даже если
используется одно из современных средств, то остается вопрос выбора
удачного начального значения.

Очевидным ответом на данный вопрос будет использование какого-либо
программного пакета, реализующего набор статистических тестов. Однако
все известные авторам пакеты реализованы на языках C или C++ и для
применения их функций непосредственно необходимо низкоуровневое
программирование. Так как язык программирования для систем
компьютерной алгебры является высокоуровневым
проблемно-ориентированным языком программирования, то внедрение
функций на языках типа C/C++ может быть невозможно или весьма
трудоёмко.

На наш взгляд это затруднение можно обойти, использовав утилиты
командной строки. Использование подобных утилит избавит от
необходимости внедрять код на языках типа C/C++ в программу и позволит
ограничиться созданием скрипта, передающего последовательность
анализируемых чисел на вход тестирующей утилите.

\section{Статистические тесты}

Тестирование генераторов псевдослучайных чисел фактически является
классической задачей проверки статистических гипотез. Но если в
математической статистике обычно ставится задача опровержения нулевой
гипотезы, то при тестировании генераторов псевдослучайных чисел,
наоборот, нулевую гипотезу пытаются подтвердить. Иными словами тесты
направлены на обоснование того, что сгенерированные последовательности
псевдослучайных чисел являются случайными независимо распределёнными
случайными величинами и не связаны никаким детерминированным
законом. Генератор успешно проходит тест, если не удалось найти
статистически значимых отклонений от ноль гипотезы.

Так как генератор псевдослучайных чисел является детерминированным
алгоритмом, то всегда найдётся тот или иной статистический тест,
который данный генератор не пройдет. Качество генератора определяется
тем, насколько много тестов он может пройти. Поэтому при создании
программной реализации тестов их объединяют в наборы тестов
и применяют совместно.

Перечислим несколько тестов, которые входят в набор статистических тестов~\texttt{DieHard}.
\begin{itemize}
\item Тест на игру в кости. Генерируется 200000 последовательностей
  равномерно распределённых псевдослучайных чисел. Каждая из
  полученных последовательностей используется для симуляции игры в
  кости. Далее проверяется, насколько теоретические значения
  математического ожидания и дисперсии согласуется с
  эмпирическими. Для статистической проверки используется критерий
  $\chi$-квадрат.
\item Парадокс дней рождения. Генерируется последовательность на
  большом интервале. Расстояния между числами должны быть
  асимптотически распределены по Пуассону.
\item Пересекающиеся перестановки. Генерируется большое число выборок
  из пяти последовательных случайных чисел. Вероятности появления
  каждой из всех возможных перестановок должны быть статистически
  эквивалентны.
\end{itemize}

Всего в DieHard входило 12 тестов. В современных программных пакетах,
реализующих наборы статистических тестов, количество
тестов существенно увеличено и превышает несколько десятков.

\section{Программные пакеты наборов статистических тестов}

Как уже отмечалось, исторически первым программным пакетом,
реализующим набор статистических тестов (или просто набор тестов) для
тестирования генераторов псевдослучайных чисел, был пакет программ
DieHard~\cite{L_DieHard}, созданный в 1995 году Джорджем Марсальей
(George Marsaglia). Он распространялся на CD-диске и в настоящее время
официальная страница доступна только в виде архива. Пакет DieHard на
данный момент не актуален, но тесты, входившие в его состав, сейчас
включены в другие программные пакеты статистических тестов.

Из актуальных в настоящее время программных пакетов, реализующих
наборы тестов, можно выделить следующие четыре.
\begin{itemize}
\item \texttt{TestU01}~\cite{L_TestU01,L_Ecuyer:2007} за авторством
  Пьера Л'Экуйе и Ричарда Симарда (Pierre L'Ecuyer, Richard
  Simard). Написана на \texttt{ANSI C}. На сегодняшний день является
  самым известным наборов тестов. Тестирует генераторы, генерирующие
  числа из интервала $[0, 1)$. Последняя версия \texttt{1.2.3} от 18
  августа 2009 года.
\item \texttt{PractRand}~\cite{L_PractRand} за авторством Криса
  Доти-Хамфри (Chris Doty-Humphrey). Написана на \texttt{С++11} с
  элементами C99. Принимает на вход поток байт, может тестировать 32 и
  64 битные генераторы. Способен справляться с очень большими объемами
  данных. Последняя версия \texttt{0.94} от 04 августа 2018 года.
\item \texttt{gjrand}~\cite{L_gjrand}. Контактов автора на официальном
  сайте найти не удалось. Написан на \texttt{С99}. Принимает на вход
  поток байт. Поставляется с набором различных генераторов, способных
  генерировать не только равномерно-распределенные последовательности
  псевдослучайных чисел, но и последовательности, подчиняющиеся
  нормальному, пуассоновскому и некоторым другим
  распределениям. Последняя версия \texttt{4.2.1} от 28 ноября 2014
  года.
\item \texttt{DieHarder}~\cite{L_DieHarder} за авторством Роберта
  Брауна. Позиционируется как наследник тестов DieHard. Написан на
  языке \texttt{C}. Требует для своей работы библиотеки
  \texttt{GSL}~\cite{L_GSL} и может тестировать любой генератор, с
  интерфейсом в стиле интерфейсов генераторов из GSL. Последняя версия
  \texttt{3.31.1} от 19 июня 2017 года.
\end{itemize}

В таблице~\ref{table:tests} дана сводка основных характеристик
обозреваемых пакетов программ. В колонке \textbf{Unix} указана возможность
установки под *nix системами. В колонке \textbf{Windows} поставлен
плюс только если программу возможно собрать без установки
\texttt{CygWin} или \texttt{MinGW}. При должном старании любую
библиотеку можно скомпилировать и под Windows тоже.

\begin{table}
  \caption{Сводная характеристика пакетов программ для тестирования}
  \label{table:tests}
  \centering
  \begin{tabular}{| c || c | c | c | c | c | c | c |}
    \hline
    \textbf{Пакет} & \textbf{Язык} & \textbf{CMD} & \textbf{Unix} & \textbf{Windows} & \textbf{Версия} & \textbf{Год} & \textbf{Сайт} \\ \hline
    \texttt{TestU01} & ANSI C & - & ++ & $\pm$ & 1.2.3 & 18.08.2009 & \cite{L_TestU01}\\ 
    \texttt{PractRand} & C99, C++11 & + & + & + & 0.94 & 04.08.2018 & \cite{L_PractRand}\\
    \texttt{gjrand} & C99 & + & + & $\pm$ & 4.2.1 & 28.11.2014 & \cite{L_gjrand}\\
    \texttt{DieHarder} & C99 & + & ++ & $\pm$ & 3.31.1 & 19.06.2017 & \cite{L_DieHarder}\\\hline
  \end{tabular}
\end{table}

Все перечисленные программные пакеты наборов тестов имеют открытый исходный код. TestU01 и
DieHarder доступны для установки через официальные репозитории многих
дистрибутивов, в частности Ubuntu 18.10. Два других набора тестов
необходимо устанавливать путем сборки из исходных кодов. Каждый из
перечисленных программных пакетов
позволяет проводить тесты как путем подключения библиотек к С/С++
программе пользователя, так и предоставляет утилиту командной
строки. Наиболее функциональная утилита командной строки у программного пакета
DieHarder.

\subsection{Установка пакетов тестов под ОС типа Unix}

Опишем процесс установки в домашний каталог пользователя без прав
администратора. Все действия выполнялись в операционной системе
GNU/Linux Ubuntu 18.10. Использовался набор компиляторов \texttt{gcc} 8
версии. Из документации к пакетам следует, что подойдет любой
компилятор, поддерживающий стандарты языков С до C99 включительно и
C++ до С++11 включительно.

В домашнем каталоге пользователя создадим следующую иерархию каталогов:
\begin{lstlisting}
  mkdir -p ~/usr/bin ~/usr/lib ~/usr/share ~/usr/include
\end{lstlisting}

\begin{itemize}
\item Каталог \path{~/usr/bin} для исполняемых файлов.
\item Каталог \path{~/usr/lib} для файлов разделяемых и статических
  библиотек (\texttt{.so} и \texttt{.a}).
\item Каталог \path{~/usr/include} для заголовочных файлов.
\item Каталог \path{~/usr/share} для примеров и документации.
\end{itemize}
Также в файл \path{~/.bashrc} добавляем следующие переменные
окружения:
\begin{lstlisting}
  export PATH="$HOME/usr/bin/:$PATH"
  export LD_LIBRARY_PATH="$HOME/usr/lib:$LD_LIBRARY_PATH"
  export LIBRARY_PATH="$HOME/usr/lib:$LIBRARY_PATH"
  export C_INCLUDE_PATH="$HOME/usr/include:$C_INCLUDE_PATH"
\end{lstlisting}
это даст возможность командному интерпретатору искать исполняемые
файлы в том числе и в каталоге \path{~/usr/bin}, а компилятору
автоматически подключать библиотеки и заголовочные файлы.

\subsubsection{Установка TestU01}

Для установки TestU01 скачаем zip архив с официального
сайта~\cite{L_TestU01}, распакуем его и перейдем в корневую
директорию.
\begin{lstlisting}
  wget http://simul.iro.umontreal.ca/testu01/TestU01.zip
  uz TestU01.zip
  cd TestU01-1.2.3/
\end{lstlisting}
TestU01 поставляется с набором скриптов \texttt{Autoconf} и
\texttt{Automake}, поэтому процесс компиляции и установки сводится к
выполнению следующих трех команд:
\begin{lstlisting}
  ./configure --prefix=$HOME/usr
  make
  make install
\end{lstlisting}
Указание опции \verb|--prefix=$HOME/usr| позволяет установить
программу локально в \path{~/usr}.

После установки в директории \path{~/usr/include} будет создано
большое количество заголовочных файлов. Для того, чтобы организовать
их аккуратней, создадим директорию \path{~/usr/include/testu01} и
перенесем туда все \texttt{.h} файлы TestU01. То же самое можно сделать
и директории \path{~/usr/lib} для файлов библиотек.

\subsubsection{Установка gjrand}

Скачиваем архив с исходными кодами с официального
сайта~\cite{L_gjrand}:
\begin{lstlisting}
  wget https://datapacket.dl.sourceforge.net/project/gjrand/gjrand/ gjrand.4.2.1/gjrand.4.2.1.tar.bz2
  tar -xvjf gjrand.4.2.1.tar.bz2
  cd gjrand.4.2.1
\end{lstlisting}
Для компиляции \texttt{gjrand} используются bash-скрипты
\texttt{compile}. Исходные файлы для получения библиотеки расположены
в директории \path{src}.
\begin{lstlisting}
  cd src && ./compile
\end{lstlisting}
на выходе получаем скомпилированный файлы динамической
\texttt{gjrand.so} и статической \texttt{gjrand.a} библиотек, которые
вручную переместим в директорию \path{~/usr/lib}:
\begin{lstlisting}
  cp -t ~/usr/lib/gjrand gjrand.a gjrand.so
  cp -t ~/usr/include/gjrand gjrand.h
\end{lstlisting}
Вернемся в корневую директорию командой \verb|cd ..| и пройдёмся
таким же образом по всем поддиректориям. В каждой из них присутствует
скрипт \texttt{compile}, который необходимо выполнить.
\begin{lstlisting}
  cd testother/src && ./compile
\end{lstlisting}
На выходе получаем исполняемые файлы в \path{testother/bin}. Вновь
вернемся на уровень выше \verb|cd ..| и перейдем в следующую
директорию:
\begin{lstlisting}
  cd testmisc/ && ./compile
\end{lstlisting}
получаем исполняемый файл \texttt{kat}, который можно использовать для
проверки корректности работы библиотеки. При запуске он выполнит ряд
тестов, чтобы проверить работает ли программа так, как рассчитывал
автор.

Перейдя в следующий каталог
\begin{lstlisting}
  cd testunif/src && ./compile
\end{lstlisting}
на выходе получим исполняемые файлы для каждого статистического
теста. Они будут располагаться в директории
\path{testunif/bin}. Непосредственно в \path{testunif} появятся два
файла \texttt{mcp} и \texttt{pmcp} --- командные утилиты для
тестирования генерируемых битовых последовательностей. Эти утилиты
используют исполняемые файлы из директории \path{testunif/bin} поэтому
переместить их в другую директорию нельзя.

Следующий каталог
\begin{lstlisting}
  cd testfunif/src && ./compile
\end{lstlisting}
полностью аналогичен предыдущему, только тесты предназначены для чисел
типа \texttt{double} из интервала $[0, 1)$.

В каталоге \path{testother}
\begin{lstlisting}
  cd testother/src && ./compile
\end{lstlisting}
находятся тесты для неравномерных распределений. Исполняемые файлы
также будут помещены в \texttt{testother/bin}.

В результате установки \texttt{gjrand} мы получили два файла
библиотеки в каталоге \path{~/usr/lib/gjrand} и заголовочный файл в
каталоге \path{~/usr/include/gjrand}. Остальные утилиты останутся в
соответствующих каталогах.

Следует отметить, что процесс установки прошел без ошибок. Автор
пакета указал опцию \texttt{-Wall} для компилятора и в процессе сборки
были видны лишь незначительные предупреждения об использовании условия
\texttt{if} без ограничивающих скобок.

\subsubsection{Установка DieHarder}

Для сборки программы необходимо наличие библиотеки GSL (GNU Scientific
Library)~\cite{L_GSL}. В Ubuntu ее можно установить выполнив команду
\begin{lstlisting}
  apt install libgsl-dev
\end{lstlisting}

Скачиваем архив с исходным кодом с официального
сайта~\cite{L_DieHarder} и распаковываем:
\begin{lstlisting}
  wget https://webhome.phy.duke.edu/~rgb/General/ dieharder/dieharder-3.31.1.tgz
  uz dieharder-3.31.1.tgz
  cd dieharder-3.31.1
\end{lstlisting}

Для установки DieHarder, как и в случае
\begin{lstlisting}
  ./configure --prefix=$HOME/usr
  make
  make install
\end{lstlisting}
Дополнительно при выполнении \texttt{./configure} можно указать опцию
\verb|--disable-shared|, что приведет к статической компиляции утилиты
командной строки и динамическая библиотека не будет создана. Это
удобно, если вам нужна только утилита командной строки.

В процессе компиляции в нашем случае произошла ошибка: компилятор не
смог найти определение типа \verb|intptr_t|. Это можно исправить,
включив в файл \path{./include/dieharder/libdieharder.h} заголовочный
файл \path{stdint.h}. Дальнейшая установка прошла без ошибок. В
процессе выполнения \verb|make install| в директорию \path{~/usr/lib}
были перенесены библиотечные файлы, в \path{~/usr/include} была
автоматически создана поддиректория \path{dieharder} и в нее помещены
все заголовочные файлы. В \path{~/usr/bin} оказалась утилита командной
строки \verb|dieharder|.

\subsubsection{Установка PractRand}

Скачиваем архив с исходным кодом с официального сайта~\cite{L_PractRand} и распаковываем:
\begin{lstlisting}
  wget https://netcologne.dl.sourceforge.net/project/ pracrand/PractRand_0.94.zip
  uz PractRand_0.94.zip
\end{lstlisting}

В архиве поставляются уже собранные файлы динамической и статической
библиотек для ОС Windows. Также присутствует файл проекта для Visual
Studio. Для установки же под Unix перейдем в директорию \path{unix}, а
сборку запустим командой \verb|make|
\begin{lstlisting}
  cd PractRand_0.94/unix
  make
\end{lstlisting}
Для сборки необходим компилятор для языка \texttt{C++}, поддерживающий
стандарт \texttt{C++11}.

При сборке возникали ошибки, связанные с регистром букв в названии
заголовочных файлов. Так, например файл \path{Coup16.h} был указан в
\path{test.cpp} в нижнем регистре, хотя само имя файла начинается с
заглавной буквы. Еще несколько таких ошибок были вызваны той же
ошибкой в именах файлов \path{NearSeq.h}, \path{birthday.h} и каталога
\path{Tests}. Такая путаница вызвана скорее всего тем, что автор
разрабатывал программу под ОС Windows, где регистр значения не имеет.

После устранения ошибок получаем 4 исполняемых файла и статическую
библиотеку \path{libpracrand.a}.

Описав установку всех четырёх рассматриваемых программных пакетов,
перейдём к описанию утилит для тестирования последовательностей
псевдослучайных чисел, которые эти пакеты предоставляют.

\subsection{Утилиты командной строки}

Пакеты \texttt{PractRand}, \texttt{DieHarder} и \texttt{gjrand} имеют
в своем составе утилиты командной строки. Данные утилиты позволяют
проводить статистические тесты над последовательностью случайных чисел
считывая ее из стандартного потока ввода или из файла. Наиболее
функциональная утилита входит в состав \texttt{DieHarder}.

\subsubsection{Утилиты PractRand}

После компиляции и сборки PractRand мы получим в свое распоряжение
четыре исполняемых файла.
\begin{itemize}
\item \verb|RNG_output| --- позволяет запустить один из генераторов,
  входящих в состав пакета.
\item \verb|RNG_test| --- утилита, предназначенная для проведения
  статистических тестов над встроенными генераторами или над потоком
  данных, подаваемом на стандартный ввод.
\item \verb|RNG_benchmark| --- измерения скорости работы встроенных
  или добавленных пользователем генераторов.
\item \verb|Test_calibration| --- для тестирования и настройки наборов
  тестов.
\end{itemize}

Для целей тестирования используется \verb|RNG_test|. Для примера,
запуск теста для встроенного генератора \texttt{jsf32} достаточно
выполнить команду
\begin{lstlisting}
  RNG_test jsf32
\end{lstlisting}
Для теста внешней программы нужно подать генерируемый бинарный поток
беззнаковых целых чисел на стандартный вход \verb|RNG_test|, например
так:
\begin{lstlisting}
  ./random -G 4 -N 1000000000000000 | RNG_test stdin32
\end{lstlisting}
Опция \texttt{stdin32} заставляет \verb|RNG_test| интерпретировать
поток бинарных данных как набор $32$-битных чисел. Если указана опция
\verb|stdin64|, то числа будут восприниматься как $64$ битные, а при
указании \verb|stdin| программа сама решит какую разрядность
использовать. Для тестирования данных из файла можно поступит,
например так:
\begin{lstlisting}
  cat file.data | RNG_test stdin64
\end{lstlisting}
Данные в файле также должны быть в бинарном формате. Данные в
текстовом формате не поддерживаются.

Тесты проводятся довольно быстро и на выход распечатывается отчет, где
перечисляются проваленные и подозрительные тесты. Указав опцию
\verb|-p 1| можно заставить программу печатать все результаты, а не
только проваленные.

\subsubsection{Утилиты gjrand}

После сборки появляется множество разных исполняемых файлов. Для
тестирования внешних генераторов предназначены два из них. Один
расположен в директории \path{testfunif} и называется \verb|mcp|
(master computer program), а второй в директории \path{testother} и
называется \verb|fmcp|. Утилита \verb|mcp| предназначена для
тестирования последовательностей целых беззнаковых псевдослучайных
чисел, а \verb|fmcp| для чисел из интервала $[0, 1)$. В остальном
между ними отличий нет.

\begin{table}
  \caption{Опции \texttt{mcp} пакета gjrand}
  \label{table:gjrand}
  \centering
  \begin{tabular}{| l | l |}
    \hline
    \textbf{Опция} & \textbf{Описание} \\ \hline
    \texttt{--tiny} & (10 MB)  \\
    \texttt{--small} & (100 MB) \\ 
    \texttt{--standard} & (1 GB) (default) \\
    \texttt{--big} & (10 GB) \\ 
    \texttt{--huge} & (100 GB)\\
    \texttt{--tera} & (1 TB) \\
    \texttt{--ten-tera} & (10 TB) \\
    \texttt{number} & количество байтов \\
    \texttt{--no-rewind} & не повторяться \\
    \hline
  \end{tabular}
\end{table}

Программа \verb|mcp| принимает на стандартный вход только бинарный
поток. Делается это с помощью конвейера
\begin{lstlisting}
  ./random -G 4 -N 1000000000000000 | mcp --big
\end{lstlisting}
К переданному потоку применяется весть набор тестов. Результаты
распечатываются по мере появления. Других опций, кроме настройки
объёма проверяемых данных (см. таблицу~\ref{table:gjrand}) у программы
нет.

\subsubsection{Утилита \texttt{dieharder}}
\label{dieharder}

После установки пакета \texttt{DieHarder} в консоли станет доступна
команда \texttt{dieharder}. Данная команда позволяет запускать и
тестировать встроенные в \texttt{DieHarder} и библиотеку \texttt{GSL}
генераторы псевдослучайных чисел. Кроме того, она может тестировать
поток случайных чисел из файлов или из стандартного ввода.  Подробную
справку по всем возможным опциям можно получить указав опцию
\texttt{-h}. Перечислим кратко наиболее важные из них:
\begin{itemize}
\item \verb|-l| --- показать список доступных тестов.
\item \verb|-d n| --- выбрать для применения тест \texttt{n}.
\item \verb|-a| --- применить все тесты.
\item \verb|-g -1| --- показать список доступных генераторов
  псевдослучайных чисел. Выбранный генератор можно протестировать
  указав опцию \verb|-g n|, где \texttt{n} --- номер генератора из
  списка.
\end{itemize}

В списке генераторов присутствуют, в частности, следующие опции.
\begin{itemize}
\item Опция \verb|200| позволяет считывать поток бинарных данных,
  подаваемый на стандартный ввод \texttt{stdin}.
\item Опции \verb|201| и \verb|202| включают считывания данных из
  файла в бинарном формате и текстовом формате соответственно.
\item Опции \verb|500| и \verb|501| позволяют считывать поток байтов с
  псевдоустройств \path{/dev/random} и \path{/dev/urandom}.
\end{itemize}

Для тестирования внешних генераторов наиболее предпочтительным
способом будет передача непрерывного потока двоичных данных через
конвейер (\verb+|+). Сделать это можно, например, следующим
способом:
\begin{lstlisting}
  ./random -G 4 -N 1000000000000000 | dieharder -g 200 -a
\end{lstlisting}
Программа \texttt{random} будет использовать генератор под номером 4
для генерации $10^{15}$ чисел. Заметим, что это не приведет к
зависанию процесса, так как \texttt{dieharder} сам закроет канал и
завершит процесс, когда проведет все тесты.

В случае считывания псевдослучайных числе из текстового файла, данный
файл должен иметь определенную структуру. Каждое псевдослучайное число
должно располагаться на новой строке, а в первых строках файла
необходимо указать следующие данные: тип чисел (\verb|d| --- целые
числа двойной точности), количество чисел в файле и разрядность чисел
($32$ или $64$ бита). Приведем пример начала такого файла:
\begin{lstlisting}
    type: d
    count: 5
    numbit: 64
    1343742658553450546
    16329942027498366702
    3111285719358198731
    2966160837142136004
    17179712607770735227
\end{lstlisting}
Когда такой файл создан можно передать его \verb|dieharder| для
проведения тестирования
\begin{lstlisting}
dieharder -a -g 202 -f file.in > file.out
\end{lstlisting}
флаг \verb|-f| задаёт входной файл с числами для анализа. Результаты
тестирования будут сохранены в \verb|file.out|. Для полноценного теста
более $10^9$ чисел, поэтому размер файла может превысить десятки
гигабайт. В случае меньшего количества чисел \texttt{dieharder} может
начать считывать файл сначала, что приведет к ухудшению результатов
теста.

\section{Пример тестирования генераторов SymPy и Maxima}

Рассмотрим как можно организовать тестирование любого генератора
псевдослучайных чисел на примере CAS Maxima и библиотеки SymPy для
Python. Будем использовать CAS Maxima версии 5.42.1 и Python 3.6.8,
дистрибутив Miniconda. Нашей целью не является сравнение реализаций
генераторов, а лишь предоставление примера, который может быть
использован для тестирования генераторов в других библиотеках,
математических пакетах и системах компьютерной алгебры.

Так как SymPy является python-модулем, то для генерации
псевдослучайных чисел используется стандартный модуль
\texttt{random}. Следующий фрагмент кода позволяет организовать вывод
беззнаковых 64-битных целых чисел в стандартный поток вывода.

\begin{lstlisting}
import random
import sys
while True:
    r = random.randint(0,2**64-1)
    r = r.to_bytes(8, byteorder="little", signed=False)
    sys.stdout.buffer.write(r)
\end{lstlisting}

В данном фрагменте генерируется целое число из полуинтервала
$[0, 2^{64})$, затем оно преобразуется в двоичный вид и выводится в
стандартный поток вывода. Записав данный скрипт в файл
\texttt{rand\_test.py} мы можем организовать тестирование выполнив в
консоли следующую команду:
\begin{lstlisting}
  python rand_test.py | dieharder -g 200 -a
\end{lstlisting}
или
\begin{lstlisting}
  python rand_test.py | RNG_test stdin64
\end{lstlisting}

В случае Maxima сгенерировать последовательность случайных целых чисел
можно с помощью следующего фрагмента кода:
\begin{lstlisting}
  for i: 1 thru 10 step 1 do print(random(2^64-1))$
\end{lstlisting}
Однако на выходе получим столбик целых чисел в текстовом виде, поэтому
для тестирования данного генератора воспользуемся возможностью утилиты
\texttt{dieharder} считывать данные в текстовом виде из заранее
подготовленных файлов (см. раздел~\ref{dieharder}).

Так как обе системы компьютерной алгебры используют алгоритм вихрь
Мерсенна, то полученные результаты тестирования практически
одинаковы. Утилита \verb|RNG_test| для SymPy выдает результат
\verb|no anomalies in 133 test result(s)|, а утилита отображает в
отчете 5 слабо пройденных тестов из более чем тридцати. Результаты
dieharder для Maxima практически идентичны.

\section{Заключение}
\label{sec:conclusion}

Авторы постарались дать исчерпывающее описание практики
построения стенда для исследования программной реализации генераторов
псевдослучайных чисел. 
Описанное программное обеспечение дает исследователю выбор из
трёх утилит командной строки. Эти утилиты снимают необходимость
низкоуровневого программирования. Кроме того, их удобно использовать
вместе с каким либо промежуточным связующим языком программирования.

Для использования этих утилит необходимо обеспечить вывод в
стандартных поток результатов генерации чисел в бинарном виде. Так как
большинство систем компьютерной алгебры ориентированны на
высокоуровневое программирование, то обеспечить бинарный вывод может
быть затруднительно. В этом случае можно использовать только утилиту
\texttt{dieharder}, так как она может обрабатывать данные в текстовом
виде.

В конце работы демонстрируются приёмы исследования программных
реализаций генераторов случайных чисел на примере двух свободных систем
компьютерной алгебры.

\def\acknowledgmentsname{Благодарности}

\begin{acknowledgments}

Публикация подготовлена при поддержке Программы РУДН «5-100».

\end{acknowledgments}

  \bibliography{rnd-cas}